\documentstyle[11pt,paspconf,epsf]{article}

\begin{document}

\title{Does merger-induced core-collapse produce $\gamma$-ray
bursts in type Ib \& Ic supernovae?}

\author{J. Middleditch}
\affil{Modeling, Algorithms, \& Informatics, CCS-3, MS B265,
Computers \& Computational Science Division,
Los Alamos National Laboratory, Los Alamos, NM 87545, jon@lanl.gov}

\begin{abstract}

Gamma-ray bursts, discovered$^1$ over three decades ago, can appear
to be a hundred times as luminous as the brightest supernovae.
However, there has been evidence for some time now of an
association$^{2,3}$ of $\gamma$-ray bursts with supernovae of type Ib and Ic.
Here we interpret the overabundance of millisecond pulsars in
globular clusters and the details of supernova 1987A to reveal
the energy source, which powers at least some long-duration
$\gamma$-ray bursts, as core-collapse following the merger$^4$ of
two white dwarfs, either as stars or stellar cores.  In order
for the beams/jets associated with $\gamma$-ray bursts to form
in mergers within massive common envelopes (as with SN1987A),
much of the intervening stellar material in the polar directions
must be cleared out by the time of core-collapse, {\it or} the beams/jets
themselves must clear their own path.  The core-collapse produces
supernovae of type Ib, Ic, or II (as with SN1987A, a SNa IIp),
leaving a weakly magnetized neutron star remnant with a spin
period near 2 milliseconds.  There is no compelling reason to
invoke any other model for $\gamma$-ray bursts.
\end{abstract}


In 1987 my colleagues and I discovered$^5$ the first pulsar in a
globular cluster, with a period of 3 milliseconds.  It soon
became clear$^6$ that there were many more millisecond pulsars
(MSPSRs) in these clusters than their supposed progenitors,
the X-ray binaries, could account for under the standard
``recycling'' theory$^7$, by about a factor of 100.  In recycling,
an ancient, solitary, slowly spinning neutron star is spun up
by accretion from a captured companion.  In spite of more than
16 years of effort, recycling has not yet successfully accounted
for the GC MSPSR population.  The most careful treatment$^8$ of this
issue confirmed that there had to be a mechanism, other than
(gradual) post-collision accretion, either from companions or
disruption disks, which could form weakly magnetized,
rapidly-spinning MSPSRs.

The only other way to get a white dwarf (WD) in a GC with a mass
exceeding the Chandrasekhar limit of 1.4 solar is to merge two
(or more) WDs$^9$.  If core-collapse via merger is possible, then
formation of MSPSRs via merger, particularly with a binary
companion, almost always dominates recycling (barring a very low
probability, resulting from the collision, for merger/core-collapse,
relative to companion-capture, evolution, accretion, and spinup).
The cross section for binary-binary collisions is always larger
than those of the binary-single, and, {\it a fortiori}, the single-single
collisions, necessary for the isolated neutron star to capture a
companion from which to eventually accrete matter.  Moreover, each
of the two stars left over from the merger process could persist
as post-SNa binary companions.

The same year also saw the SN1987A outburst, followed shortly by
the discovery$^{10,11}$ of the ``mystery spot'' (now a pair$^{12}$ on opposite
sides of, and in line with, the axisymmetric$^{13}$ ejecta), ~0.06 arc s
(17 light days in projection) south of SN1987A, and with a luminosity
nearly 5\% of maximum light. Like the overabundance of MSPSRs in
the GC's, this spot has never been reconciled with traditional models 
(of SNe).

When the structure and kinematics of the three rings surrounding SN1987A
started to become clear almost a decade ago$^{14,15}$, other colleagues
proposed a binary merger scenario$^{16}$.  In this picture, the inner ring
was formed by mass loss through at least one of the two outer mass axis
Lagrangean points, efficiently producing its extremely low (as compared
to blue supergiant winds) 10 km/s expansion velocity, essentially the
thermal velocity of hydrogen at photospheric temperatures$^{17}$.  The polar
gradients of the potential just beyond L2 and L3, may have helped
collimate the gas outflow to the observed small ring angular height.
Neither member of the binary was likely to have had a recent red
supergiant wind$^{18}$ due to the limitations of space in the close binary.
 
Extrapolating this line of reasoning, the two fainter, outer rings around
SN1987A were likely formed close to the epoch of the contact binary during
which the inner ring was formed.  Radiation pressure from one or both
stars on their companion's outer atmosphere may have produced the gentle
wind of ~25 km/s, least disturbed by the orbital motion for the two
directions, nearest to, but at least 30 degrees from, their respective
poles$^{19}$.  The actual half angle of the cones, which perhaps could be used
to determine the pre-merger mass ratio, is a more realistic 48 degrees,
with the flow, over the whole domain of polar angles, forming the observed
fans of extended emission$^{15}$.

Later, the nearly merged star was likely concave in both polar directions,
and still later convex, the transition possibly causing polar ejection of
the material that we can speculate formed the target into which a polar
jet/beam dumped$^{11}$ (an isotropic) $10^{49}$ ergs {\it en passant}.  To be conservative,
the spot is assumed {\it not} to be material traveling at relativistic$^{12}$ speed,
prior to being hit by the jet/beam, which {\it was} relativistic.

Later yet, prior to core-collapse, angular momentum transfer from the two
merging WD cores might have at least partially cleared out the star's
inner polar regions. Having the SNa-associated beam/jet itself blast
through, or carry along, the normally intervening stellar material, seems
even more improbable. The pre-SNa clearing process, if sufficiently abrupt,
may have spectrophotometric consequences, which could be exploited to form
an ``early warning system'' for merger SNe in most of the BSG's in the
Local Group and blue stragglers in the globulars.

The merged WD would rotate$^{16}$ with a period near 1.98 s, set by the
branching between Jakobi and Maclaurin configurations, and, if its mass
exceeded 1.4 solar, core-collapse would follow in many$^{4}$ or most cases.
A neutron star/pulsar with a spin period near 2 ms would form within a
type Ib, Ic, or II SNa, consistent, in the case of the IIp SN1987A, with
the 2.14 ms signal$^{20}$.

It has already been suggested that the mystery spot in SN1987A, about
24 light days distant from the pre-explosion binary's pole, was a result
of a lateral GRB$^{21}$.  But the crucial link to understanding GRBs may be
that some or most type Ib, Ic, \& II SNe, which may comprise more than 90\%
of all SNe, are the result of WD-WD mergers. Since, as we have argued,
SN1987A was the result of such a merger, and certainly produced a neutron
star, then we know that such mergers can indeed produce neutron star
remnants, which can also be MSPSR's$^{20}$.  The vast overabundance of the
MSPRS in the GC's strongly supports this assertion.

The recent discovery of five transient X-ray MSPSRs in the Galactic plane$^{22}$
does not challenge this view.  The same mechanism, which likely produces
most of the binary and solitary MSPSRs in the globulars, works at least
as well as recycling (see above), even in the Galactic plane.  In fact,
further discoveries of such sources, with none still in the GCs, becomes
embarrassing to the application of recycling everywhere, all the while not
even exclusively supporting recycling in the Plane.

Unlike most type Ib and Ic SNe, SN1987A entrained and/or encountered a
much higher density of material, thus what radio emission that was
generated may have been self-absorbed.  It is also possible that the
radio lobes themselves are beamed. Beaming, in general, removes the
$10^{54}$ erg energy requirement for GRB's that ``collapsars'' or ``hypernovae,''
massive stars which collapse to black holes, were invented to satisfy$^{23,24}$.
The same reasoning holds for the vast majority of type Ib and Ic SNe
which show no evidence of the central engine$^{25}$ as was associated$^{26}$ with
SN1998bw, near GRB980425, with the additional caveat that perhaps not all
type Ib and Ic SNe result in a neutron star remnant. The yield can not be
so low, however, as to deprive the GC's of their quota of MSPRSs.  In
addition, core-merger for type Ib and Ic SNe doesn't suffer from the
difficulty, in producing the apparent maximum energy observed in the radio
lobes$^{27}$, that plagues these theories.
 
The extremely high expansion velocities$^{23}$ seen in SNe associated with
GRB's may be due to very early observations, never made of others
because their discovery takes more time, or to variations in the mass
of the common envelope.  Additional energetic events following the
initial burst could just be due to other gas targets farther away in the
polar directions of the binary merger. Finally, magnetars can be
constrained observationally during the SNa decline, and later (as in$^{28}$
Cas A) by the expected X-ray modulation of the still hot neutron star,
via residual accretion or magnetic-thermal interaction.

The nature of the energy source behind GRB's can {\it definitively} be
determined only, perhaps, for pulsars, from magnetars to MSPSRs,
by detecting an underlying periodicity. Thus, taking ``imposter'' SNe
into account, high time-resolution observations of all nearby ($< 50$ Mpc)
SNe should be made, using the largest and most sensitive optical and radio
telescopes, to search for pulsar remnants. Opacity will not be an issue for
many of these.  Observations of only SNe which happen to be associated with
GRB's might not be fruitful, as these are usually much farther away, and
the proximity of the line of sight to the rotation axis of the binary
merger could reduce the amplitude of the pulsar signal. A number of
extragalactic SNe (though no Ib's or Ic's) have already been searched$^{29}$,
and this has continued sporadically, without success, though only scant
resources have been devoted to this effort so far. Continuing indefinitely
without further, more sensitive searches is to risk wasting time and
resources. Taking the results from SN1987A at face value$^{20}$ means that the
pulsed optical signal was 10 or more solar luminosities, at 5.0 -- 6.5 years
of age.  This is hardly surprising, as the Crab pulsar's (nearly
entirely pulsed) optical output is 4 solar, and nanosecond radio bursts
in pulsars are bright enough to be seen in neighboring galaxies$^{30}$.

Occam's razor alone should have been sufficient to restrict models to
the realm of the feasible/already observed phenomena, had we only paid
more attention to observations of the GC's and SN1987A. This new
understanding of GRB's may represent the end of an era, begun four decades
ago, of exploration and discovery in high-energy astrophysics.

\acknowledgements

I thank Kaiyou Chen, Stirling Colgate and  Chris Fryer for 
their support, guidance, conversations and  insight, and 
Alexander Heger for timely comments.  The work  was performed 
under the auspices of the Department of Energy.

\end{document}